\documentclass[a4paper,11pt]{article}
\usepackage{jinstpub} 
\usepackage{lineno}
\usepackage{multirow}
\usepackage{graphicx}
\usepackage{lipsum}
\usepackage{natbib}
\usepackage[english]{babel} 

\usepackage{pgfplots}
\pgfplotsset{compat=1.18}

\usepackage[inkscapelatex=false]{svg}

\title{\boldmath Readout chain of SPD DAQ system}

\author[a]{L. Afanasyev}
\author[b]{A. Berngardt}
\author[a]{A. Boikov}
\author[b]{V. Borshch}
\author[b]{D. Erofeev}
\author[a]{V. Frolov}
\author[a]{K. Gritsay}
\author[a]{V. Olshevsky}
\author[a]{A. Kulikov}
\author[a]{S. Tereshchenko}
\author[a]{V. Tereshchenko}

\affiliation[a]{Joint Institute for Nuclear Research, Dubna, Russia}
\affiliation[b]{The National Research Tomsk State University, Tomsk, Russia}

\emailAdd{boikov@jinr.ru}
\emailAdd{aleksandr.boikov@cern.ch}

\abstract{The Spin Physics Detector (SPD) experiment at the NICA collider of the Joint Institute for Nuclear Research aims to study the spin structure of the nucleon and spin-related phenomena. The experimental conditions, including the high interaction rate and the large number of readout channels, require a triggerless data acquisition system with high throughput and a scalable architecture.

This work presents a hardware implementation of the readout chain that extends the previously proposed SPD DAQ architecture. The chain is based on two levels of concentrators. The L1 concentrator provides the interface between the front-end electronics and the next system level and also distributes synchronous signals. The L2 concentrator aggregates data streams from several L1 concentrators before transmitting them to the computing system. A key feature of the L2 implementation is the offloading of data reception, buffering, and aggregation from the host CPU to the FPGA, motivated by the large number of small packets generated by the readout electronics.

The developed hardware platforms and their operating logic are presented, together with experimental results for the main interfaces and data-path components. The obtained results demonstrate the feasibility of the proposed hardware architecture and its application in a scalable SPD data acquisition system.}


\begin{document}
\maketitle
 \flushbottom

\section{Introduction}

The Spin Physics Detector (SPD) experiment at the NICA collider aims to study the spin structure of the nucleon and spin-related phenomena in interactions of polarized proton and deuteron beams. The detector comprises several subdetectors based on different detection principles, including measurements of signal time and amplitude. In the full configuration, the number of detector channels may reach about $6\times10^5$, while the expected total data rate is estimated at 20~GB/s~\cite{TDR}.

The high interaction rate, the proximity of background processes to physics signals, and the need to reconstruct event kinematics substantially limit the applicability of a conventional hardware trigger system. Therefore, the SPD experiment intends to use a triggerless data acquisition (DAQ) system in which data from the front-end electronics (FEE) continuously enter the DAQ system without preliminary hardware event selection~\cite{SPD-DAQ-Concept}. This approach preserves information about rare physics processes that can only be selected at subsequent stages of data processing.

Operation of the DAQ system in this mode imposes stringent requirements on the data transmission and processing infrastructure. The system must provide:
\begin{itemize}
    \item continuous transmission of large data volumes from all detector channels;
    \item precise time synchronization of distributed elements;
    \item scalability with increasing numbers of readout channels;
    \item control of the FEE and monitoring of its status.
\end{itemize}

The general DAQ architecture for the SPD experiment was proposed in~\cite{SPD-DAQ-Concept}, including a hierarchical structure of concentrators, data-flow organization principles, and a White Rabbit-based time-synchronization system~\cite{WhiteRabbitSpecification}. The present work extends this study and focuses on the hardware implementation of the key elements of the proposed architecture. Particular attention is paid to the development of first-level (L1) and second-level (L2) concentrators, high-speed interfaces to the FEE, and experimental characterization of their performance. The aim of this work is to demonstrate the operability of the proposed hardware solutions and evaluate how they correspond to DAQ requirements in terms of throughput, latency and scalability.
\section{Architecture}

The readout chain of the SPD DAQ system is organized into two hierarchical levels for receiving, aggregating and transmitting data from the FEE to the computing system. Such organization also allows the system to scale as the number of detector channels increases.

\begin{figure}
\centering
\includegraphics[width=1\textwidth, keepaspectratio]{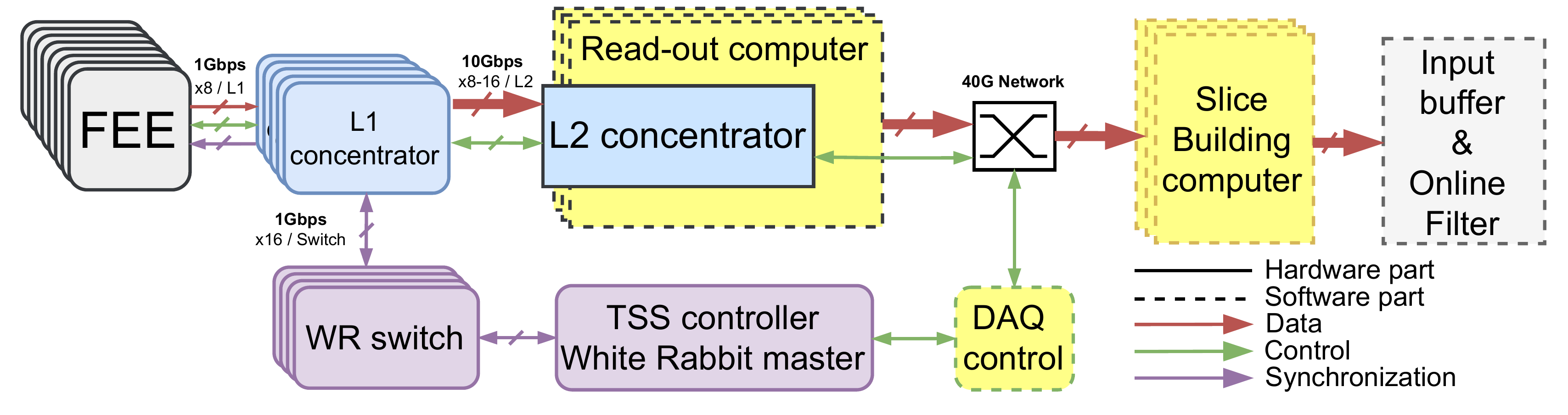}
\caption{General structure of the SPD data acquisition system: hierarchy of L1 and L2 concentrators, data flows from the FEE to the computing system, and distribution of White Rabbit synchronization signals.}
\label{DAQ-struct}
\end{figure}

As shown in figure~\ref{DAQ-struct}, data from the FEE are transmitted to L1 concentrators, where several input streams are combined into one. The aggregated data are then transmitted to L2 concentrators, which combine streams from several L1 concentrators and forward them to the computing system through a PCIe interface using direct memory access (DMA).

In parallel with the data path, the system uses a White Rabbit-based time-synchronization system (TSS) ~\cite{WhiteRabbitSpecification} to distribute the global clock and synchronous commands. The end nodes of this network are the L1 concentrators located close to the FEE. This provides a common time reference for the data, enabling their organization into time slices. In the SPD architecture, data are organized into a sequence of slices grouped into frames; each slice is uniquely identified by a frame number and its number within the frame~\cite{SPD-DAQ-Concept}.

Since the FEE generates a large number of short packets, their processing creates a significant load on the network stack of the computing system. To reduce this load, L2 implements a slice-block formation algorithm that reduces the communication overhead between L2 and the computing system and performs throughput-critical operations directly in the field-programmable gate array (FPGA). The details of this data flow and its impact on performance are discussed in the section devoted to L2.

One L1 concentrator provides connections for up to eight FEE boards, while one L2 concentrator can combine up to 16 L1 concentrators. The minimum number of concentrators at each level is determined by the total number of FEE boards, $N_{\mathrm{FEE}}$:

\begin{equation}
N_{\mathrm{L1}} =
\left\lceil
\frac{N_{\mathrm{FEE}}}{8}
\right\rceil ,
\qquad
N_{\mathrm{L2}} =
\left\lceil
\frac{N_{\mathrm{L1}}}{16}
\right\rceil .
\end{equation}

The expected number of FEE-L1 links for different detector configurations is given in table~\ref{tab:l1-l2}. The detector is planned to be implemented in two phases. In Phase 1, the initial configuration, 1,292 links require at least 162 L1 concentrators and 11 L2 concentrators. In Phase 2, the number of concentrators depends on the vertex-detector configuration: 4,811 FEE-L1 links for Double-Sided Silicon Detectors (DSSD) require 602 L1 and 38 L2 concentrators, while 5,800 FEE-L1 links for Monolithic Active Pixel Sensors (MAPS) require 725 L1 and 46 L2 concentrators. The choice between DSSD and MAPS for the vertex detector will be made at a later stage of the experiment.

\begin{table}[ht]
\centering
\caption{Minimum number of L1 and L2 concentrators for different detector configurations.}
\label{tab:l1-l2}
\begin{tabular}{lccc}
\hline
\small
\textbf{Configuration} & \textbf{FEE boards} &
\textbf{L1} & \textbf{L2} \\ \hline
Phase 1       & 1\,292 & 162 & 11 \\
Phase 2, DSSD & 4\,811 & 602 & 38 \\
Phase 2, MAPS & 5\,800 & 725 & 46 \\ \hline
\end{tabular}
\end{table}

The values above define the minimum number of concentrators. The number of FEE boards connected to individual L1 concentrators may vary depending on the detector geometry and FEE placement, which should be as close as possible to the detector elements. Therefore, the final system may contain additional concentrators to provide a throughput margin and balanced loading.

Providing sufficient throughput margin at all points of the readout chain from FEE to L2 is one of the key principles of the system design. This principle, introduced in the original DAQ architecture~\cite{SPD-DAQ-Concept}, helps to avoid bottlenecks and ensures continuous data transmission. The hardware implementation is based on commercially available components with long-term support; the use of standard interfaces and FPGA technology provides reproducibility and adaptability to changes in the detector configuration.

Thus, the proposed architecture provides scalability of the DAQ system: the transition from the first phase of the experiment to the full configuration is achieved by adding L1 and L2 concentrators without changing the principles of the hierarchy or the interfaces between its levels.
\section{L1 Concentrator}

The L1 concentrator is an intermediate node of the readout chain that connects the FEE to the L2 concentrator. L1 combines data streams from the connected FEE boards and transmits them to L2, while also distributing time-synchronization signals.

The L1 concentrator is based on an Intel Cyclone 10 GX FPGA, which implements data processing and transmission, FEE interface control, interaction with the synchronization system, and network communication with L2. The basic configuration provides connections for eight FEE boards, with the possibility of upgrading to a 16-port configuration.

Data transmission from L1 to L2 uses an optical interface with a data rate of up to 10~Gbit/s. A separate optical channel with a data rate of 1~Gbit/s is used for communication with the White Rabbit time-synchronization system. The L1 hardware contains the clock generation units required to operate as a White Rabbit node in accordance with the corresponding specification~\cite{WhiteRabbitSpecification}.

The L1 design provides several implementation options depending on the installation location. For installation directly within the detector volume, a compact version can be used, optimized for size and connection-line length. For installation in standard electronics racks outside the detector volume, a VME-crate module version is provided.

\begin{figure}
\centering
\includegraphics[width=0.6\textwidth, keepaspectratio]{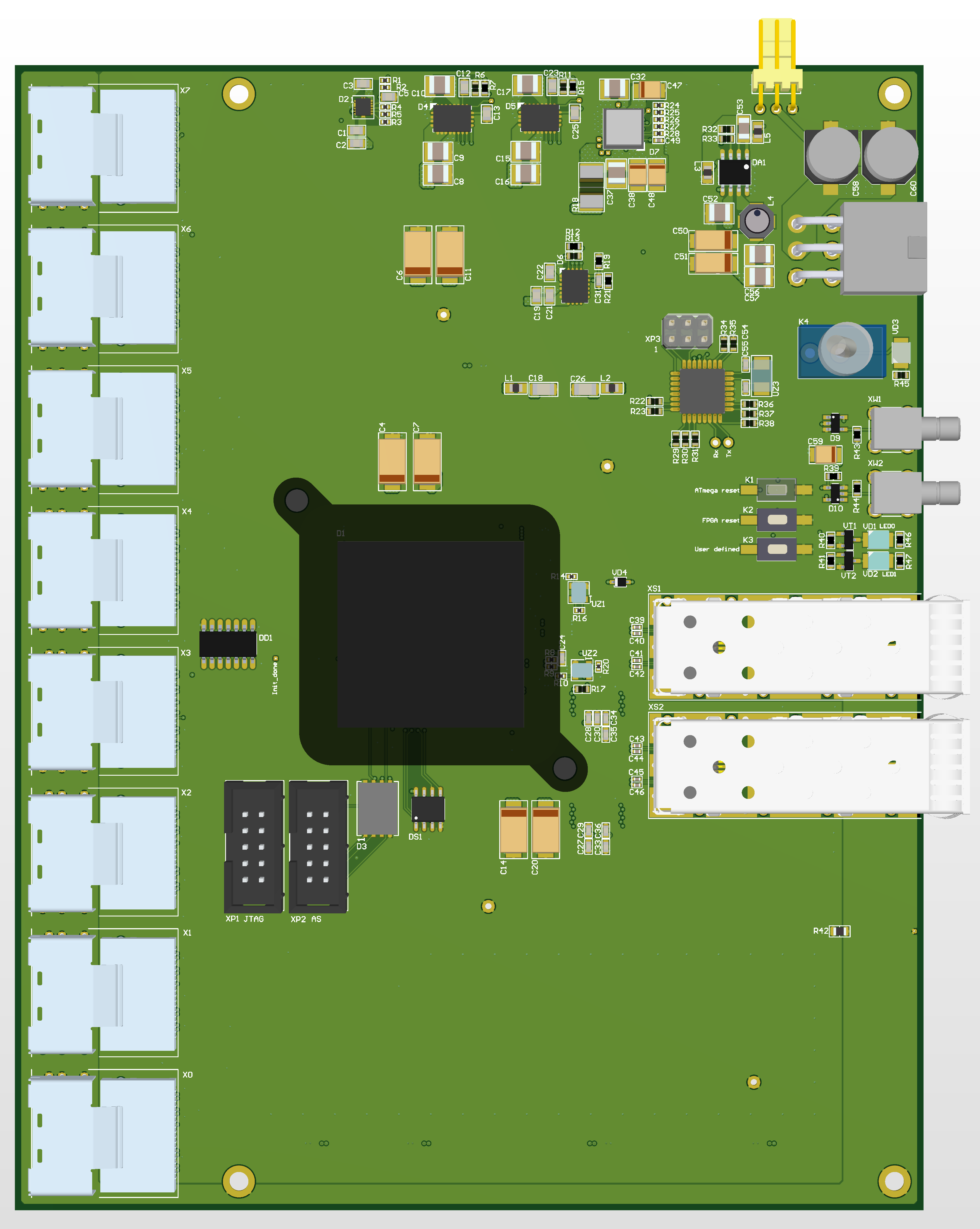}
\caption{L1 concentrator printed circuit board.}
\label{l1-hub-PCB}
\end{figure}

\subsection{L1--FEE interface}

Communication between the L1 concentrator and the FEE is implemented using a copper differential interface. The developed version uses a MiniSAS-based cable system that provides the required number of high-speed and auxiliary lines in a single connection. A similar approach is used in other experiments; for example, the MiniSAS interface is used to connect L1DDC and MMFE8 boards in the ATLAS New Small Wheel electronics~\cite{ATLAS-NSW-L1DDC}.

The interface supports bidirectional communication between L1 and the FEE. In the L1--FEE direction, packet-based control commands are transmitted for FEE configuration and control, including access to internal registers, status monitoring, and remote updating of the FPGA configuration memory. In the opposite direction, the FEE transmits generated data and responses to the control commands.

For each FEE in the L1--FEE direction, two differential lines are used: one for data transmission and one for the clock signal. Before transmission, the data are encoded using 8b/10b coding and serialized. On the FEE side, the received stream is deserialized and decoded.

In the FEE-L1 direction, one differential line carries the data. No separate clock line is used; the clock is recovered directly from the received stream using clock and data recovery (CDR). This reduces the number of physical interface lines.

In addition to the high-speed data lines, the interface contains separate lines for distributing synchronization signals and implementing auxiliary functions. The FEE uses Start of Slice (SoS) and Start of Frame (SoF) signals together with a synchronous 125~MHz clock for synchronization. Additional single-ended signals are used for hardware interaction between the FEE and L1, FEE reset, and a backup slow-control interface.

\subsection{Synchronization signal distribution}

A Time Synchronization system (TSS) node is implemented in the concentrator FPGA. It interfaces with the White Rabbit core, receives timing information, recovers the synchronous clock, and generates synchronous control commands.

The commands are transmitted from L1 to the FEE through dedicated differential lines. The propagation delay of the L1--FEE connection is compensated in the TSS node so that synchronous commands reach the connected FEE boards with consistent timing.

The SoS and SoF signals define the time structure of the data stream, while the synchronous 125~MHz clock serves as the reference for the FEE synchronous logic. These signals are distributed together so that the connected FEE boards operate with a common system time base in accordance with the SPD DAQ architecture~\cite{SPD-DAQ-Concept,TDR}.

\subsection{Experimental Characterization of the L1--FEE Interface}

The operation of the L1--FEE electrical interface was experimentally investigated at a line rate of 1.25~Gbit/s. Generation and reception of the serial stream were performed using the SERDES (serializer/deserializer) blocks of the Intel Cyclone 10 GX FPGA. With 8b/10b encoding, the specified line rate corresponds to a useful data rate of 1~Gbit/s. The test stream used the K28.5 symbol for symbol-boundary detection and receiver alignment. The use of K28.5 for alignment of 8b/10b symbol boundaries is a standard approach for serial streams using this coding scheme~\cite{cit_8b10b}.

\begin{figure}
\centering
\includegraphics[width=1\linewidth]{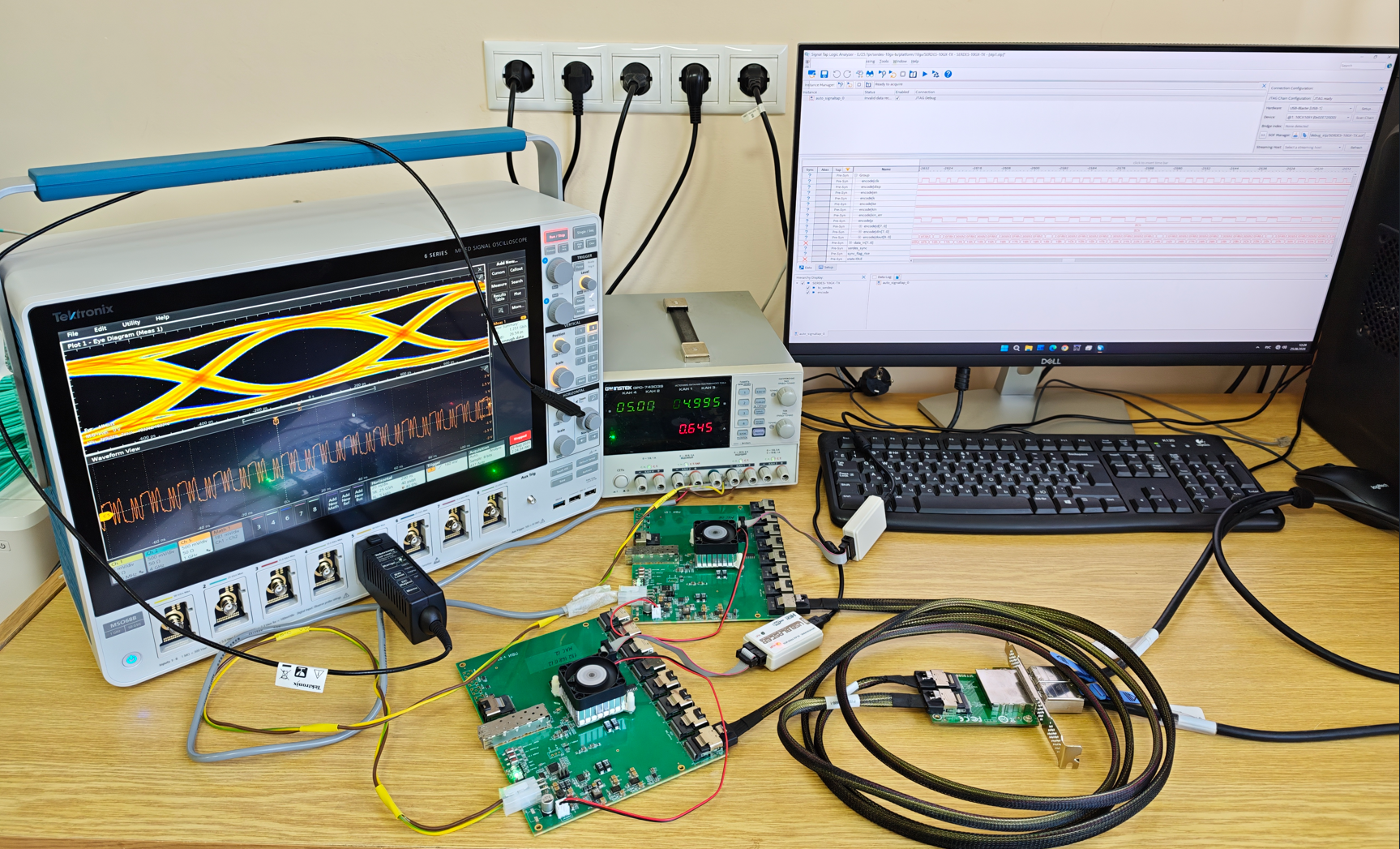}
\caption{Test stand for investigation of the L1--FEE electrical interface.}
\label{MinisasTB}
\end{figure}

\begin{figure}
\centering
\includegraphics[width=1\linewidth]{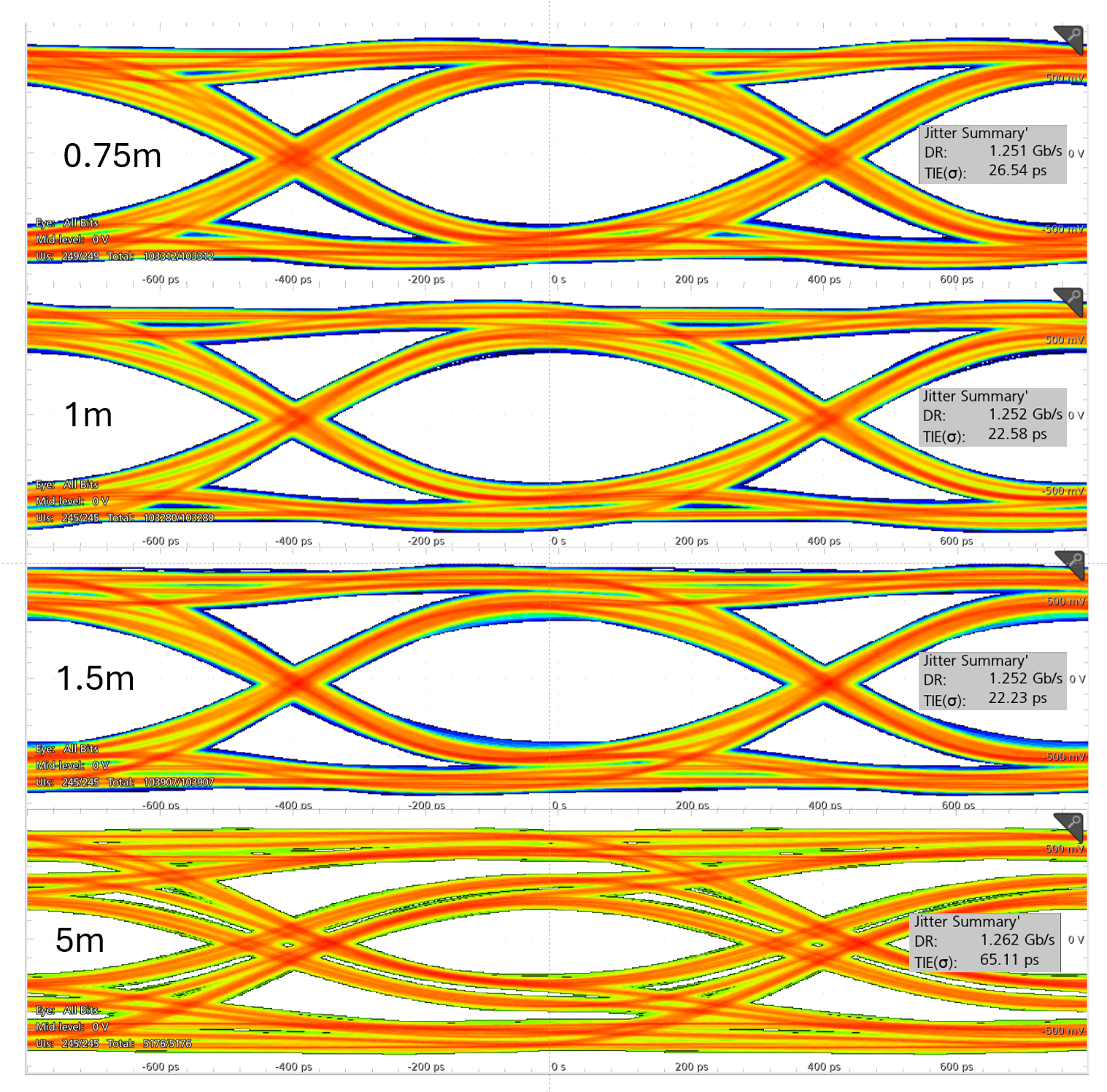}
\caption{Eye diagrams of the received signal for different cable lengths.}
\label{eye_minisas}
\end{figure}

To evaluate the interface for different cable lengths, cables with lengths of 0.75~m, 1~m, and 1.5~m were investigated. In addition, a 5~m cable assembly was tested. It consisted of two 1.5~m MiniSAS cables, a 2~m SAS cable, and a passive adapter connecting the cable sections. This configuration was used to evaluate the interface with a long cable assembly and to investigate the influence of additional connections and physical-path discontinuities. Transmission quality was evaluated using eye diagrams of the received signal~\cite{LVDS_Cable_Performance}. Measurements were performed with a Tektronix MSO 68b oscilloscope and a TDP1000 differential probe. The test stand is shown in figure~\ref{MinisasTB}, and the measurement results are presented in figure~\ref{eye_minisas}.

When testing the 0.75~m-long cable, the measured time interval error (TIE) was 26~ps, while the eye diagram showed a clear eye opening in both amplitude and time. When the connection length was increased to 1--1.5~m, the TIE was 22~ps, with no significant change in the eye opening. For the 5~m cable assembly, the TIE increased to 65~ps and the eye opening decreased substantially. Nevertheless, no errors in reception of the test stream were observed under the conditions of the experiment.

Thus, the L1--FEE electrical interface remains operational for a cable assembly with a total length of 5~m at a line rate of 1.25~Gbit/s. At the same time, the measured eye diagram and increased TIE indicate that this configuration is close to the operating limit of the interface. For the 5-m line, the TIE is approximately 2.5--3.2 times higher than the values obtained for the other samples, accompanied by a noticeable reduction in the eye opening.

The reduced eye opening observed for the 5~m cable assembly indicates increased signal distortion. Attenuation of high-frequency components slows the signal transitions, while impedance discontinuities at the cable interfaces can cause reflections and further distort the waveform. As a result, the received signal becomes more sensitive to the transmitted bit sequence, reducing the eye opening and the signal-integrity margin. These effects explain the increase in TIE observed for the 5~m cable assembly.

\subsection{L1 Firmware}

\begin{figure}
\centering
\includegraphics[width=1\linewidth]{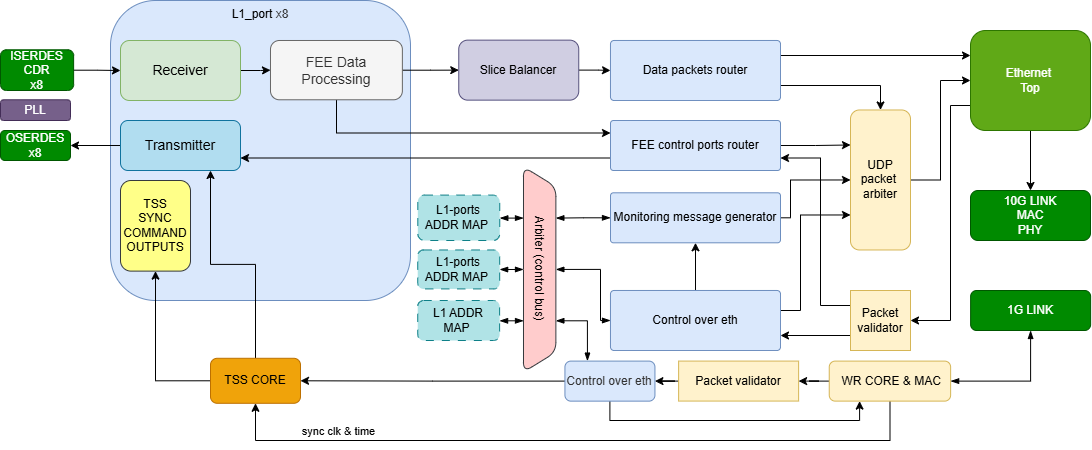}
\caption{Simplified functional structure of the L1 concentrator firmware.}
\label{l1-hub-Firmware}
\end{figure}

The internal structure of the L1 concentrator firmware is shown in figure~\ref{l1-hub-Firmware}. The firmware is logically divided into functional blocks responsible for communication with the FEE, data processing and routing, data transmission to L2, control and status monitoring, and time-synchronization functions.

The L1-ports block provides the physical communication channels with the FEE. It includes data reception and transmission logic, stream serialization and deserialization, encoding and decoding, clock recovery, and link-state control. The block architecture scales to the required number of connected FEE boards according to the hardware configuration of the concentrator.

Data received from the FEE are passed to the Slice Balancer block, which provides sequential transmission of slices to the next system element. After the Slice Balancer, the data are passed to the Data Packet Router, which performs routing and forms network packets for transmission to L2.

The network block provides communication between L1 and L2 using the UDP/IP/ICMP stack. Its main purpose is to transmit the formed data stream to L2. The same interface provides access to internal L1 and FEE registers, transmission of control and service commands, and collection and transmission of device-status information.

The synchronization-system interface block includes the TSS node and the White Rabbit core. It enables the L1 concentrator to operate as a White Rabbit network node and transmit synchronous commands to the connected FEE boards.

Thus, the L1 implementation integrates the FEE hardware interfaces, data processing and routing, control functions, and time synchronization in a single FPGA device.
\section{L2 Concentrator}

The second-level concentrator (L2) combines data streams from several first-level concentrators (L1) and transmits the aggregated data to the Readout computer. Unlike L1, which is directly connected to the FEE, L2 performs the aggregation of multiple input streams in hardware and forms the data stream for the computing system.

In triggerless mode, the FEE generates a large number of relatively small packets. If these packets are transmitted directly to the computing system, the processing load depends not only on the total data rate but also on the number of individual packet-processing operations on the Readout computer side. A high packet rate increases the load on the network subsystem and reduces the efficiency of the available computing resources.

L2 therefore performs the aggregation directly in the FPGA. Packets received from several L1 concentrators are buffered and combined into larger data blocks according to the \textit{slice-block} structure. The resulting blocks are written to DDR4 memory placed on the L2 board and transferred to the memory of the Readout computer through a PCI Express interface using direct memory access (DMA).

This organization shifts data reception, buffering, and aggregation from the software network subsystem to the FPGA hardware. The Readout computer consequently receives larger data blocks instead of a large number of individual packets. L1 performs the initial combination of FEE streams, while L2 carries out their further aggregation before transmission to the computing system.

Each L2 is connected to several L1 concentrators through independent high-speed optical channels. The architecture can be scaled by increasing the number of L1 and L2 concentrators without changing the principles used to build individual DAQ branches.

The main tasks of the L2 concentrator are:
\begin{enumerate}
\item reception and aggregation of data streams from several L1 concentrators;
\item preliminary checking and routing of incoming packets;
\item structuring data by FEE and time slice (Slice);
\item buffering data in external DDR4 memory;
\item formation of data sets corresponding to individual time slices;
\item generation of descriptors for data transmission using Scatter-Gather DMA;
\item transmission of aggregated data to the Readout computer via PCI Express;
\item routing of control commands and monitoring information between the control system, L1, and FEE.
\end{enumerate}

\subsection{Hardware implementation}

At the initial stage of development, the Z19-P evaluation platform based on the Xilinx Zynq UltraScale+ FPGA was used for L2 prototyping~\cite{l2_concentrator_2024,l2_conentrator_2025}. The platform was used to verify the L2 architecture, individual functional blocks, and the required FPGA resources.

Based on the prototyping results, the target hardware platform was selected based on an FPGA from the Xilinx Kintex UltraScale+ family implemented as a System-on-Module (SOM). The selected SOM is equipped with 10~GB of DDR4 SDRAM, as well as 32 GTH and 24 GTY transceivers. These resources provide the required high-speed L2 interfaces, including optical links to L1 and a PCI Express Gen3 x16 interface.

Unlike the Xilinx Zynq UltraScale+ FPGA used at the prototyping stage, the selected FPGA does not contain an integrated processor subsystem. Therefore, the DDR4 memory is directly accessible to the programmable logic (PL), allowing it to be used for hardware buffering and data-flow organization without sharing it with a processor subsystem.

\begin{figure}[ht]
\centering
\includegraphics[width=1\textwidth,keepaspectratio]{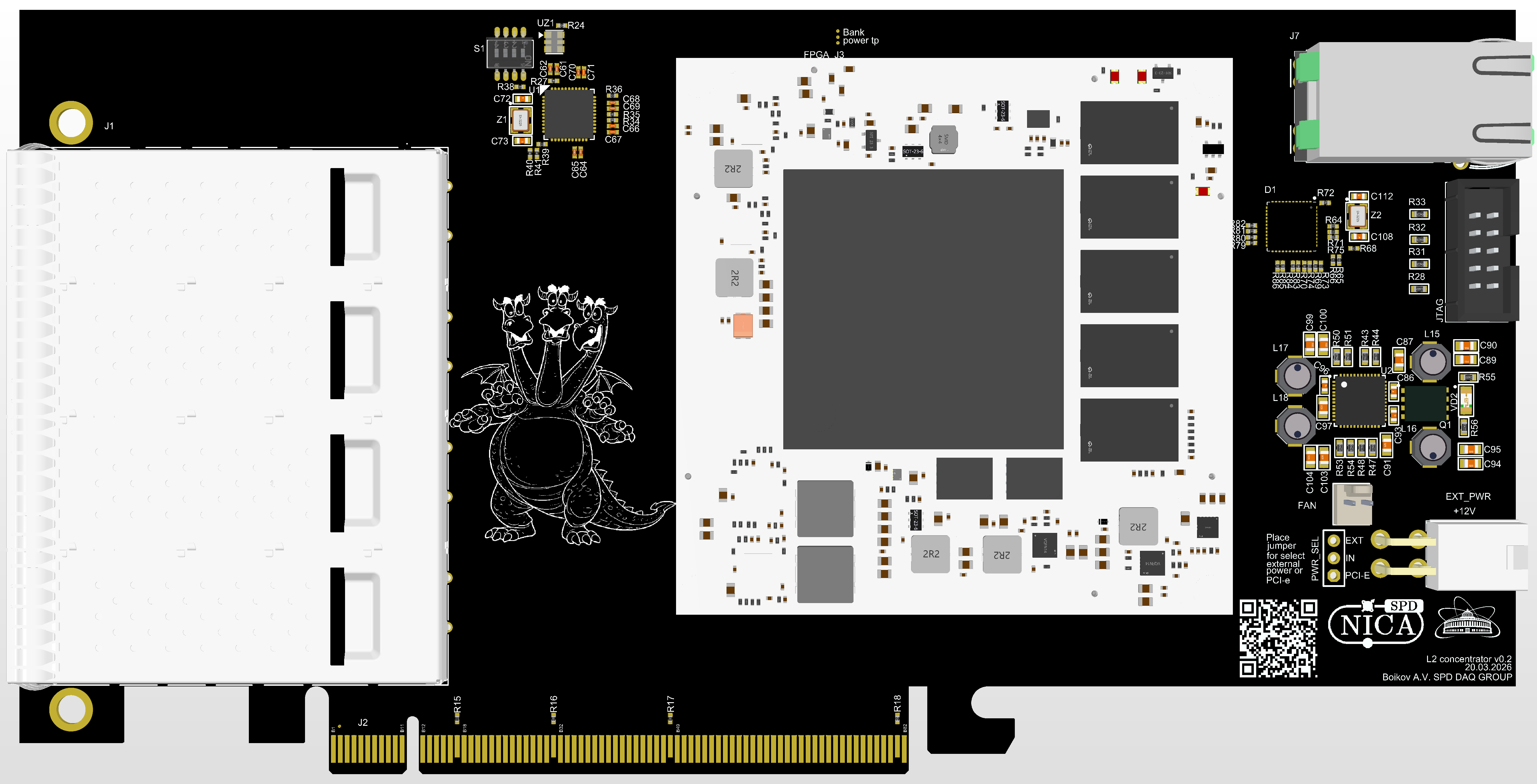}
\caption{Prototype of the L2 hardware platform.}
\label{l2-hub-PCB}
\end{figure}

The SOM is installed on a dedicated carrier board (figure~\ref{l2-hub-PCB}) that provides external interfaces and L2 peripherals. The board contains a PCI Express connector for connection to the Readout computer, four QSFP connectors for optical channels, and circuits for generation and distribution of reference clock signals for the transceivers. Sixteen independent optical data-transmission channels are implemented for communication with L1, with an Ethernet rate of 10~Gb/s per channel. The serial physical line rate is 10.3125~Gb/s per channel, corresponding to a total physical line rate of 165~Gb/s for 16 channels. The physical interfaces are grouped into four channels in each of four QSFP modules.

In addition to the high-speed interfaces, the carrier board contains a power-generation and distribution subsystem for the SOM and peripheral components, as well as auxiliary circuits required for FPGA configuration, programming, and debugging.

Thus, each L2 provides connections for up to 16 L1 concentrators. At the Ethernet level, the total nominal data rate is 160~Gb/s. The effective Ethernet payload rate additionally depends on the transmitted frame size because part of the transmission time is occupied by Ethernet overhead fields and the inter-frame gap.

\subsection{Data transmission efficiency between L1 and L2}

To evaluate transmission efficiency, Ethernet overhead must be taken into account. The total amount of data occupying the transmission time of one frame on the physical channel is defined as

\begin{equation}
L_{\mathrm{line}} =
L_{\mathrm{payload}} +
L_{\mathrm{MAC}} +
L_{\mathrm{FCS}} +
L_{\mathrm{preamble}} +
L_{\mathrm{IFG}},
\end{equation}
where $L_{\mathrm{payload}}$ is the size of the Ethernet-frame payload,
$L_{\mathrm{MAC}}=14$~B is the Media Access Control (MAC) header size,
$L_{\mathrm{FCS}}=4$~B is the frame check sequence (FCS) size,
$L_{\mathrm{preamble}}=8$~B is the preamble and start frame delimiter (SFD),
and $L_{\mathrm{IFG}}=12$~B is the inter-frame gap (IFG).

Thus, the total fixed overhead is 38~B. For one channel with an Ethernet rate of 10~Gb/s, the effective payload data rate is given by

\begin{equation}
R_{\mathrm{payload}} =
10~\mathrm{Gb/s}
\frac{L_{\mathrm{payload}}}
{L_{\mathrm{payload}}+38~\mathrm{B}} .
\end{equation}

For the maximum Ethernet payload size, $L_{\mathrm{payload}}=1500$~B, the effective rate is approximately 9.75~Gb/s, or 1.22~GB/s. For the minimum payload size, $L_{\mathrm{payload}}=46$~B, it decreases to 5.48~Gb/s, or 0.69~GB/s.

For 16 independent channels, the total effective payload rate varies from 87.6~Gb/s (11.0~GB/s) at the minimum payload size to 156.0~Gb/s (19.5~GB/s) at the maximum Ethernet payload size.

For payload sizes below 46~B, the Ethernet standard requires the payload field to be padded to the minimum allowed size. Taking this into account, the effective payload rate for 16 channels is described by

\begin{equation}
R_{\mathrm{payload}}(L_{\mathrm{payload}}) =
\left\{
\begin{array}{ll}
160~\mathrm{Gb/s}\,
\displaystyle\frac{L_{\mathrm{payload}}}{84~\mathrm{B}},
&
1~\mathrm{B}\leq L_{\mathrm{payload}}<46~\mathrm{B},
\\[8pt]
160~\mathrm{Gb/s}\,
\displaystyle\frac{L_{\mathrm{payload}}}
{L_{\mathrm{payload}}+38~\mathrm{B}},
&
L_{\mathrm{payload}}\geq46~\mathrm{B}.
\end{array}
\right.
\end{equation}

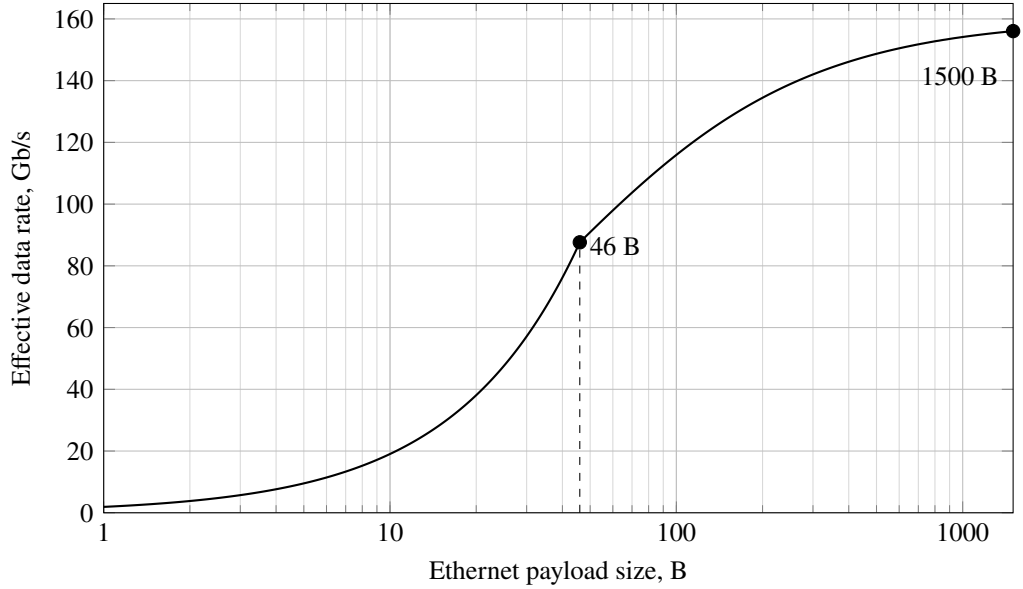
\begin{figure}[ht]
\centering
\begin{tikzpicture}
\begin{axis}[
width=0.9\textwidth,
height=0.55\textwidth,
xmode=log,
xlabel={Ethernet payload size, B},
ylabel={Effective data rate, Gb/s},
xmin=1,
xmax=1500,
ymin=0,
ymax=165,
xtick={1,10,100,1000},
xticklabels={1,10,100,1000},
grid=both,
minor grid style={gray!30},
tick label style={font=\small},
label style={font=\small},
samples=200,
]

    \addplot[
        thick,
        domain=1:46
    ] {
        160*x/84
    };

    \addplot[
        thick,
        domain=46:1500
    ] {
        160*x/(x+38)
    };

    \addplot[
        only marks,
        mark=*,
        mark size=2.5pt
    ] coordinates {
        (46,87.62)
        (1500,156.05)
    };

    \addplot[
        dashed
    ] coordinates {
        (46,0)
        (46,87.62)
    };

    \node[
        anchor=south west,
        font=\small
    ] at (axis cs:46,80)
    {$46$~B};

    \node[
        anchor=south east,
        font=\small
    ] at (axis cs:1450,135)
    {$1500$~B};

\end{axis}
\end{tikzpicture}

\caption{
    Effective payload data rate as a function of Ethernet payload size
    for 16 channels with an Ethernet rate of 10~Gb/s.
}
\label{fig:l2-payload-efficiency}
\end{figure}

The dependence of the effective payload rate on the Ethernet payload size for all 16 channels is shown in figure~\ref{fig:l2-payload-efficiency}.

After an Ethernet frame is received by L2, the Ethernet service fields are removed and subsequent processing is performed on the received payload. Therefore, the above estimate characterizes the efficiency of physical-channel utilization for data transmission between L1 and L2, but does not determine the performance of the subsequent L2--Readout computer path.

The obtained dependence shows that as the payload size decreases, the relative fraction of overhead fields and the inter-frame gap increases. As a result, the effective payload rate decreases significantly. In addition to the loss of effective bandwidth, transmission of a large number of small frames increases the number of individual packet-processing operations. This further increases the requirements for the L2 hardware logic and is one of the reasons for implementing data aggregation directly in the FPGA.

\subsection{Data transmission to the Readout computer}

Data transmission from L2 to the computing system is performed using a PCI Express Gen3 x16 interface. The interface is implemented using the XDMA IP core, which provides direct data transfer between the FPGA and the memory of the Readout computer using direct memory access (DMA).

The theoretical bandwidth of PCI Express Gen3 x16 is about 16~GB/s in each direction~\cite{Xilinx_PG213}. The actual transfer rate is lower than the theoretical value because of PCI Express protocol overhead and the characteristics of the particular DMA implementation. As an external reference, measurements for the Xilinx Alveo platform have demonstrated a host-bound FPGA data-transfer rate of about 12~GB/s~\cite{AMD_XDMA_Gen3x16}.

DMA allows the blocks formed in L2 to be transferred directly to the memory of the Readout computer without processing individual packets by the central processing unit (CPU). Thus, most data-transfer operations are performed by the hardware DMA controller, reducing CPU load.

L2 uses Scatter-Gather DMA, which allows data located in several separate DDR4 memory regions to be transferred as part of a sequence of DMA operations. This approach matches the structure of the generated \textit{slice-block}, since data from one time slice may be located in several external-memory regions during reception and sorting.

With Scatter-Gather DMA, the addresses and sizes of the corresponding memory regions are specified by a set of descriptors. Based on these descriptors, XDMA transfers the prepared DDR4 regions through PCI Express to the memory of the Readout computer. The descriptors are generated by the hardware SG Descriptor Generator after the formation of the data associated with the corresponding Slice is completed.

This approach allows the formed blocks to be transferred as soon as they are ready and eliminates the need for additional data copying inside L2. As a result, the number of data-movement operations between memory regions is reduced, providing efficient transmission of aggregated data to the computing system.

\subsection{Firmware architecture}

\begin{figure}[ht]
\centering
\includegraphics[width=1\linewidth]{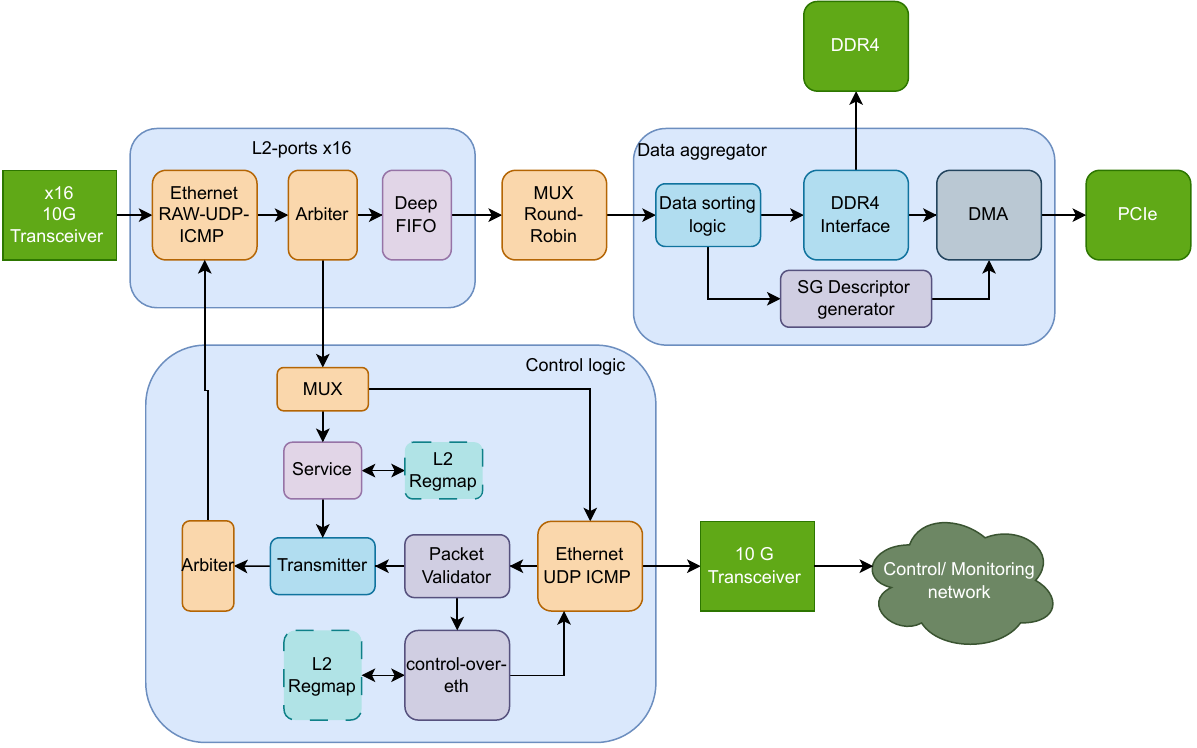}
\caption{Simplified functional structure of the L2 firmware.}
\label{l2-hub-Firmware}
\end{figure}

The functional structure of the L2 firmware is shown in figure~\ref{l2-hub-Firmware}. The architecture is divided into three main functional blocks: the block for receiving streams from L1, the block for aggregating and transmitting data to the Readout computer, and the block for routing control commands and monitoring information.

The L2-ports block receives data from 16 independent optical channels with a line rate of 10.3125~Gb/s per channel. Incoming Ethernet frames are processed by the Ethernet RAW-UDP-ICMP block, which determines the type of transmitted information and directs the data stream to the corresponding path. Data packets are passed to FIFOs, then combined by a Round-Robin arbiter and sent to the data-aggregation block.

The Data sorting logic block places incoming data according to their association with the FEE and time slice. Based on this information, \textit{slice-blocks} are formed and written to external DDR4 memory through the corresponding interface. Thus, sorting and time organization of the data are performed directly in the FPGA before transmission to the Readout computer.

After reception of the data belonging to one Slice is completed, the SG Descriptor Generator forms a set of descriptors containing the addresses and sizes of the corresponding DDR4 regions. These descriptors are used by DMA to transfer the formed data set through PCI Express to the Readout computer.

Separation of the reception and aggregation paths decouples processing of high-speed input streams from writing to DDR4 and subsequent transmission through PCI Express. External-memory buffering provides data accumulation and smooths differences between the input rate from L1 and the transfer rate to the computing system.

\subsection{Control and monitoring}

A separate Control Logic block handles control commands, configuration data, and monitoring information exchanged between the external control network and the L2, L1, and FEE nodes. The control path is physically separated from the main data path, so service operations do not directly affect high-speed data transmission.

An incoming UDP packet contains the destination address, operation type, and required parameters. Incoming packets are processed by the Ethernet UDP/ICMP and Packet Validator blocks. After validation, the packet is passed to an arbiter, which determines the required service path and directs the command either to the internal L2 control mechanisms or to the corresponding communication channel with L1.

L2 routes commands between the external control network and the connected L1 concentrators. L1, in turn, provides further transmission of commands to the corresponding FEE. Response data follow the reverse path through L1 and L2 and are transmitted to the external control network.

L2 therefore combines hardware data aggregation with control-information routing. The high-speed data path and the service path are implemented independently, separating the requirements for data transmission from those of device servicing.

\subsection{L2 summary}

The developed L2 combines up to 16 input streams from L1, each with an Ethernet rate of 10~Gb/s. The physical line rate of each optical channel is 10.3125~Gb/s. Hardware aggregation is performed in the FPGA using external DDR4 memory, after which the formed \textit{slice-blocks} are transmitted to the Readout computer via PCI Express Gen3 x16 and Scatter-Gather DMA.

This organization moves data reception, buffering, sorting, and aggregation operations into the hardware part of the system and reduces the load on the software network subsystem of the Readout computer. At the same time, independent high-speed communication channels between L1 and L2 and the possibility of scaling the number of L2 concentrators allow the architecture to meet the requirements for extending the SPD DAQ system as the number of detector channels increases.
\section{Summary and conclusions}

This work has presented a hardware implementation of the key elements of the data acquisition system for the SPD experiment, extending the previously proposed DAQ architecture~\cite{TDR, SPD-DAQ-Concept}. The main focus was on the first- and second-level concentrators, L1 and L2, and their interfaces to the FEE, the synchronization system, and the computing system.

The L1 concentrator receives data from the FEE, transmits to FEE the configuration and control commands and distributes synchronization signals. L2 aggregates data streams from several L1 concentrators and transfers them to the memory of the Readout computer. The large number of relatively small packets in the input stream increases the load on the network subsystem of the computing node. For this reason, data reception, buffering, and aggregation are implemented directly in the FPGA, reducing the number of packet-processing operations on the computing system.

The hardware solutions use commercially available components and standard high-speed interfaces. Experimental studies confirmed the operation of the developed components and the main data-transmission interfaces.

Together, L1 and L2 form the hardware data path from the FEE to the computing system, providing data transmission, synchronization, buffering, and aggregation. The modular design of the concentrators allows additional L1 and L2 units to be added as the number of readout channels and the corresponding data-processing requirements of the SPD experiment increase.
\section{Acknowledgments}
This work was partially supported by Ministry of Science and Higher Education of the Russian Federation (project No. FSWM-2025-0023) in the part of study of the L2 concentrator for readout chain.

\bibliographystyle{unsrt}
\bibliography{references}

\end{document}